\documentclass[aps,prc,superscriptaddress,nofootinbib,showkeys]{revtex4-2}

\usepackage[utf8]{inputenc} 
\usepackage[T1]{fontenc} 
\usepackage{lmodern}
\usepackage{graphicx}
\usepackage{xcolor}
\usepackage{amsmath}
\usepackage{dcolumn}
\usepackage{bm}
\usepackage{braket}
\usepackage[hidelinks]{hyperref}
\newcommand{\bmsi}{{\bm \sigma}}
\newcommand{\bmta}{{\bm \tau}}

\begin{document}
	
	
\title{Partial-wave decomposition of isospin-projected subleading three-nucleon contact interactions}
	
	
	
	
\author{Elena~Filandri}
\email{efilandri@ectstar.eu}
\affiliation{European Centre for Theoretical Studies in Nuclear Physics and Related Areas (ECT$^{*}$) and Fondazione Bruno Kessler, Villa Tambosi, Villazzano, Trento, I-38123, Italy}
	
\author{Luca~Girlanda}
\email{girlanda@le.infn.it}
\author{Ylenia~Capitani}
\email{ylenia.capitani@unisalento.it}
\affiliation{Department of Mathematics and Physics ``E.~De~Giorgi'', University of Salento, Via per Arnesano, I-73100 Lecce, Italy}
\affiliation{Istituto Nazionale di Fisica Nucleare (INFN), Sezione di Lecce, I-73100 Lecce, Italy}
	
\date{\today}
	
\begin{abstract}
We analyze the subleading three-nucleon contact interaction terms at N4LO in chiral effective field theory. 
We perform the isospin projection into $T = 1/2$ and $T = 3/2$ states, and find that only eleven of the thirteen operators contribute in the $T=1/2$ channel. By projection on the partial waves of the asymptotic configuration of $N-d$ scattering states in momentum space, we find contributions from ten combinations of LECs.
These results provide a useful basis for including N4LO three-body forces in few-body calculations and for constraining them through numerical fits to data.
\end{abstract}
	
	
	\maketitle
	

\section{Introduction}

The accurate description of few-nucleon systems constitutes a longstanding and fundamental problem in theoretical nuclear physics. The complexity of nuclear interactions, particularly in systems involving more than two nucleons, requires theoretical frameworks that are not only systematically improvable but also rooted in the underlying symmetries of quantum chromodynamics (QCD). In recent decades, Chiral Effective Field Theory ($\chi$EFT)~\cite{Weinberg1990,Weinberg1991,Weinberg:1992yk,vanKolck1999,Epelbaum:2005pn,Epelbaum2009,Machleidt:2011zz} has become the standard low-energy effective theory for nuclear forces, providing a controlled expansion in powers of momenta and pion masses relative to the breakdown scale of the theory.

A key strength of $\chi$EFT lies in its ability to incorporate two-, three-, and higher-body forces in a hierarchical fashion. At next-to-next-to-leading order (N2LO), three-nucleon (3N) forces appear naturally \cite{Epelbaum:2002vt} and are essential for explaining binding energies and scattering observables in few-body systems. As one proceeds to higher orders in the chiral expansion, subleading corrections to the 3N force emerge, introducing new operator structures and low-energy constants (LECs) that must be constrained by data.

At next-to-next-to-next-to-leading order (N3LO), the standard wisdom suggests that no new 3N contact terms arise \cite{Bernard:2007sp,Bernard:2011zr} (see however Refs.~\cite{Girlanda:2020pqn,Filandri2023,Girlanda:2024rkl} for a different point of view). Subleading 3N contact interactions contribute at N4LO via a rich operator basis involving two derivatives and more complex spin-isospin structures~\cite{Girlanda_2011,*PhysRevC.102.019903}. These subleading contributions are not only crucial for quantitative improvements in nuclear observables but also offer the possibility of resolving persistent discrepancies between theory and experiment.

One such discrepancy is the so-called nucleon analyzing power puzzle ($A_y$ puzzle) in nucleon-deuteron ($N-d$) scattering \cite{Kievsky:1995uk,Golak:2014ksa,Viviani:2013wra}. Despite the inclusion of N2LO and even N3LO 3N forces and high-order two-nucleon interactions, theoretical predictions systematically underestimate the vector analyzing power in $p-d$ and $n-d$ elastic scattering at low energies~\cite{Filandri2023, Golak:2014ksa,LENPIC:2015qsz}. Among the subleading operators introduced at N4LO, spin-orbit type terms---such as those represented by the operators $O_7$ and $O_8$---have been proposed as promising candidates to mitigate this longstanding discrepancy \cite{Kievsky:1999nw}.

In this work, starting with the general  structure of the N4LO three-nucleon contact interaction in $\chi$EFT, we identify the combinations of LECs more sensitive to the different partial waves of elastic $N-d$ scattering, with the aim of providing a useful guide to the fit procedures of experimental data. 

The paper is organized as follows.
Starting from the most general form of the subleading 3N contact potential in momentum space, composed by a set of 13 linearly independent operators consistent with all the relevant symmetries, in Section~\ref{sec:isospinproj}, we carry out the isospin projection of the operators into the total isospin $T=\frac{1}{2}$ and $T=\frac{3}{2}$ sectors.
Then, in Section~\ref{sec:pwd}, we perform a detailed partial-wave decomposition of the potential matrix elements in the $N-d$ scattering channels. 
Finally, Section~\ref{sec:conclusions} contains a summary of our main results and a discussion of their implications for the implementation of subleading 3N forces in $\chi$EFT-based few-body computations.

\section{Three-body contact interactions}\label{sec:3Bcontact}
The subleading contact contributions to the three-nucleon force have been classified in Ref.~\cite{Girlanda_2011,*PhysRevC.102.019903} as consisting of 13 independent operators with two derivatives, respecting all the discrete symmetries and requiring Poincar\'{e} invariance,
\begin{equation}
	V = - \sum_{i=1}^{13}E_i\,O_i\,,\label{eq:1}
\end{equation}
whose explicit form in momentum space is given as follows
\begin{eqnarray}
	V=\sum_{i \neq j \neq k} \big[
	&-&E_1 \mathbf{k}_i^2
	-E_2 \mathbf{k}_i^2 \boldsymbol{\tau}_i \cdot \boldsymbol{\tau}_j \nonumber\\
	&-&E_3 \mathbf{k}_i^2 \boldsymbol{\sigma}_i \cdot \boldsymbol{\sigma}_j-E_4\mathbf{k}_i^2 \boldsymbol{\sigma}_i \cdot \boldsymbol{\sigma}_j \boldsymbol{\tau}_i \cdot \boldsymbol{\tau}_j\nonumber\\
	&-&E_5\left(3 \mathbf{k}_i \cdot \boldsymbol{\sigma}_i \mathbf{k}_i \cdot \boldsymbol{\sigma}_j-\mathbf{k}_i^2 \boldsymbol{\sigma}_i \cdot \boldsymbol{\sigma}_j\right)
	-E_6\left(3 \mathbf{k}_i \cdot \boldsymbol{\sigma}_i \mathbf{k}_i \cdot \boldsymbol{\sigma}_j-\mathbf{k}_i^2 \boldsymbol{\sigma}_i \cdot \boldsymbol{\sigma}_j\right) \boldsymbol{\tau}_i \cdot \boldsymbol{\tau}_j\nonumber\\
	&-&\frac{i}{2} E_7\mathbf{k}_i \times\left(\mathbf{Q}_i-\mathbf{Q}_j\right) \cdot\left(\boldsymbol{\sigma}_i+\boldsymbol{\sigma}_j\right)
	-\frac{i}{2}E_8 \mathbf{k}_i \times\left(\mathbf{Q}_i-\mathbf{Q}_j\right) \cdot\left(\boldsymbol{\sigma}_i+\boldsymbol{\sigma}_j\right) \boldsymbol{\tau}_j \cdot \boldsymbol{\tau}_k \nonumber\\
	&-&E_9 \mathbf{k}_i \cdot \boldsymbol{\sigma}_i \mathbf{k}_j \cdot \boldsymbol{\sigma}_j
	-E_{10}\mathbf{k}_i \cdot \boldsymbol{\sigma}_i \mathbf{k}_j \cdot \boldsymbol{\sigma}_j \boldsymbol{\tau}_i \cdot \boldsymbol{\tau}_j\nonumber \\
	&-&E_{11} \mathbf{k}_i \cdot \boldsymbol{\sigma}_j \mathbf{k}_j \cdot \boldsymbol{\sigma}_i
	-E_{12}\mathbf{k}_i \cdot \boldsymbol{\sigma}_j \mathbf{k}_j \cdot \boldsymbol{\sigma}_i \boldsymbol{\tau}_i \cdot \boldsymbol{\tau}_j\nonumber\\
	&-& E_{13} \mathbf{k}_i \cdot \boldsymbol{\sigma}_j \mathbf{k}_j \cdot \boldsymbol{\sigma}_i \boldsymbol{\tau}_i \cdot \boldsymbol{\tau}_k \big] \,,
	\label{eq:1a}
\end{eqnarray}
with $\mathbf{k}_i=\mathbf{p}_i-\mathbf{p}'_i$, $\mathbf{Q}_i=\frac{\mathbf{p}_i+\mathbf{p}_i'}{2}$.
Particularly interesting terms for examining $A_y$ puzzles are given by $O_7$ and $O_8$, since, as suggested in Ref.~\cite{Kievsky_1999}, these contributions being of spin-orbit type turn out to be propitious for such a problem.
In this section, we first examine the projection on the possible three-nucleon system isospin channels for this potential; then, we perform the partial-wave decomposition.

\subsection{Isospin Projection}\label{sec:isospinproj}
It is possible to project the potential in Eq.~(\ref{eq:1}) in the two possible isospin channels, for a three-nucleon system, $T=\frac{1}{2}$ and $T=\frac{3}{2}$ . 
To identify linear combinations purely  $T=\frac{1}{2}$ and $T=\frac{3}{2}$ the following projectors are needed,
\begin{align}
	P_{1/2}&=\frac{1}{2}-\frac{1}{6}\left(\bmta_{1} \cdot \bmta_{2}+\bmta_{1} \cdot \bmta_{3}+\bmta_{2} \cdot \bmta_{3}\right)\,, \\
	P_{3/2}&=1-P_{1/2}\,.
\end{align}
Defining the fundamental isospin structures $T^+$ and $T^-$,  even and odd, respectively, under time reversal symmetry,
\begin{align}
	T^+&=\bm{1}, \bmta_i\cdot \bmta_j \,,\\
	T^-&=\bmta_i\times\bmta_j\times\bmta_k \,,
\end{align}
one can firstly calculate  how the projector $ P_{1/2}$ acts on these minimal blocks,
\begin{align}
	P_{1/2}T^+&=T^+ -1+P_{\frac{1}{2}}=T^+-P_{\frac{3}{2}}\,,\\
	P_{1/2}T^-&=T^-\,.
\end{align}
The projections of the subleading contact operators $O_{i}$ $(i=1, \ldots, 13)$ into the $T=\frac{1}{2}$ subspace of the three particles, $\left(O_{i}\right)_{1 / 2}=P_{1 / 2} O_{i} P_{1 / 2}$, read as follows (see also Ref.~\cite{Nasoni_thesis}):
\begin{align}
	\left(O_{i}\right)_{1 / 2}&=O_{i}\,, \qquad i=3, \ldots, 6 \,,
	\\
	\left(O_{1}\right)_{1 / 2} &=O_{1}-\frac{1}{3} O_{2}+\frac{1}{3} O_{3}+\frac{1}{9} O_{4}+\frac{1}{3} O_{5}+\frac{1}{9} O_{6}
	+\frac{1}{2} O_{9}+\frac{1}{6} O_{10}+\frac{1}{2} O_{11}+\frac{1}{6} O_{12} \,,
	\\
	\left(O_{2}\right)_{1 / 2} &=\frac{2}{3} O_{2}+\frac{1}{3} O_{3}+\frac{1}{9} O_{4}+\frac{1}{3} O_{5}+\frac{1}{9} O_{6}+\frac{1}{2} O_{9}
	+\frac{1}{6} O_{10}+\frac{1}{2} O_{11}+\frac{1}{6} O_{12} \,,
	\\
	\left(O_{7}\right)_{1 / 2} &=O_{7}-\frac{1}{6} O_{2}+\frac{1}{6} O_{3}+\frac{1}{18} O_{4}+\frac{1}{6} O_{5}+\frac{1}{18} O_{6}
	+\frac{1}{4} O_{9}+\frac{1}{12} O_{10}-\frac{1}{4} O_{11}-\frac{1}{12} O_{12}-\frac{1}{3} O_{13} \,,
	\\
	\left(O_{8}\right)_{1 / 2} &=O_{8}-\frac{1}{6} O_{2}+\frac{1}{6} O_{3}+\frac{1}{18} O_{4}+\frac{1}{6} O_{5}+\frac{1}{18} O_{6}
	+\frac{1}{4} O_{9} +\frac{1}{12} O_{10}-\frac{1}{4} O_{11}-\frac{1}{12} O_{12}-\frac{1}{3} O_{13} \,, \\
	\left(O_{9}\right)_{1 / 2} &=\frac{1}{2} O_{9}+\frac{1}{3} O_{2}-\frac{1}{3} O_{3}-\frac{1}{9} O_{4}-\frac{1}{3} O_{5}
	-\frac{1}{9} O_{6}-\frac{1}{6} O_{10} +\frac{1}{3} O_{13} \,, \\
	\left(O_{10}\right)_{1 / 2} &=\frac{5}{6} O_{10}+\frac{1}{3} O_{2}-\frac{1}{3} O_{3}-\frac{1}{9} O_{4}-\frac{1}{3} O_{5}
	-\frac{1}{9} O_{6}-\frac{1}{2} O_{9}+\frac{1}{3} O_{13} \,, \\
	\left(O_{11}\right)_{1 / 2} &=\frac{1}{2} O_{11}-\frac{1}{6} O_{12}-\frac{1}{3} O_{13} \,, \\
	\left(O_{12}\right)_{1 / 2} &=-\frac{1}{2} O_{11}+\frac{5}{6} O_{12}-\frac{1}{3} O_{13} \,, \\
	\left(O_{13}\right)_{1 / 2} &=-\frac{1}{2} O_{11}-\frac{1}{6} O_{12}+\frac{2}{3} O_{13} \,,
\end{align}
where use has been made of all the Fierz relations among the operators.
The purely $T=\frac{1}{2}$ and $T=\frac{3}{2}$ combinations can be found  by imposing the conditions 
$P_{1/2} c_{j}^{(i)} O_{j}=c_{j}^{(i)} O_{j}$ and $P_{1/2} c_{j}^{(i)} O_{j}=0$, with $i,j=1,\ldots,13$, where the $c_{j}^{(i)}$ are the coefficients of the linear combinations and the sum on repeated indices is omitted.

Purely $T=\frac{1}{2}$ operators result,
\begin{align}
	&\tilde{O}_{1}^{(1 / 2)}=O_{1}-2 O_{7}+O_{11}+O_{12}, \\
	&\tilde{O}_{2}^{(1 / 2)}=O_{2}-O_{1}, \\
	&\tilde{O}_{3}^{(1 / 2)}=O_{3}, \\
	&\tilde{O}_{4}^{(1 / 2)}=O_{4}, \\
	&\tilde{O}_{5}^{(1 / 2)}=O_{5}, \\
	&\tilde{O}_{6}^{(1 / 2)}=O_{6}, \\
	&\tilde{O}_{7}^{(1 / 2)}=O_{7}+\frac{1}{2} O_{9}-\frac{1}{4} O_{11}-\frac{1}{12} O_{12}-\frac{1}{6} O_{13}, \\
	&\tilde{O}_{8}^{(1 / 2)}=O_{8}-O_{7}, \\
	&\tilde{O}_{9}^{(1 / 2)}=O_{9}-O_{10}, \\
	&\tilde{O}_{10}^{(1 / 2)}=O_{11}-O_{12}, \\
	&\tilde{O}_{11}^{(1 / 2)}=O_{12}-O_{13},
\end{align}
and  purely $T=\frac{3}{2}$ operators,
\begin{align}
	&\tilde{O}_{1}^{(3 / 2)}=O_{11}+\frac{1}{3} O_{12}+\frac{2}{3} O_{13}, \\
	&\tilde{O}_{2}^{(3 / 2)}=O_{2}-O_{3}-\frac{1}{3} O_{4}-O_{5}-\frac{1}{3} O_{6}-\frac{3}{2} O_{9}-\frac{1}{2} O_{10}+O_{13} \,.
\end{align}
By equating
\begin{equation}
	\sum_{i=1}^{13} E_{i} O_{i}=\sum_{i=1}^{11} E_{i}^{(1 / 2)} O_{i}^{(1 / 2)}+\sum_{i=1}^{2} E_{i}^{(3 / 2)} O_{i}^{(3 / 2)},
\end{equation}
one has the appropriate LECs redefinition. 
Thus only 11 linearly independent combinations of LECs $E_1,\dots,E_{13}$ contribute to $p-d$ observables, in the isospin limit.

\subsection{Partial-wave decomposition}\label{sec:pwd}
We consider the scattering states of the $N-d$ system, $| \Psi_{Nd} \rangle$.
They are eigenstates of the total angular momentum $\mathbf{J}$, the total isospin  $\mathbf{T}$ and their $z$-projections, and therefore will be denoted as
\begin{align}
	| \Psi_{Nd }\rangle = | \Psi_{LST}^{JJ_zT_z}\rangle \,,
\end{align}
with $L$ the quantum number relative to the orbital momentum between the incident nucleon and the deuteron, and $S$ the total spin. 
These states are also eigenstates of the parity operator with eigenvalue $(-1)^L$.

Denoting as $\mathbf{p}_1,\mathbf{p}_2,\mathbf{p}_3$ the momenta of the three particles, we define the intrinsic momenta $\mathbf{k}^{(\pi)},\mathbf{p}^{(\pi)}$ in the center-of-mass reference frame, where $\mathbf{P}=\mathbf{p}_1+\mathbf{p}_2+\mathbf{p}_3=0$, as follows
\begin{align}
	&\mathbf{k}^{(\pi)} = \frac{1}{2}(\mathbf{p}_{\pi_2} - \mathbf{p}_{\pi_1}) \,,\label{eq:kp_k}\\
	&\mathbf{p}^{(\pi)} = \mathbf{p}_{\pi_3} \,.\label{eq:kp_p}
\end{align}
The definition depends on the permutation $\pi$, which is characterized by the group of indices $\pi=(\pi_1,\pi_2,\pi_3)$.
The ``even'' permutations correspond to $\pi=1,2,3$, with associated sign $\varepsilon_\pi=1$, while the ``odd'' ones are $\pi=4,5,6$, with $\varepsilon_\pi=-1$.
Since the scattering states are totally antisymmetric, they can be written as
\begin{equation}
	| \Psi_{LST}^{JJ_zT_z} \rangle
	= \frac{1}{\sqrt{6}} \sum_{\pi=1}^6 \varepsilon_\pi | \Psi_{LST}^{JJ_zT_z} \rangle_{\pi} \,.
\end{equation}
We construct the wave function in momentum space in the following form%
\footnote{%
	This corresponds to considering the Fourier transform of the asymptotic wave function defined in coordinate space.
}:
\begin{align}
	\Psi_{LST}^{JJ_zT_z}
	&= \tilde{\delta}(\mathbf{P}) \frac{1}{\sqrt{6}} \sum_{\pi=1}^6 \varepsilon_\pi
	\frac{\tilde{\delta}(p^{(\pi)}-q)}{p^{(\pi)2}} 
	\sum_{\ell_d=0,2} \varphi_{\ell_d}(k^{(\pi)}) \nonumber\\
	&\times
	\bigg[ Y_L(\mathbf{\hat{p}}^{(\pi)}) \Big[  \big[
	Y_{\ell_d}(\mathbf{\hat{k}}^{(\pi)}) \; \chi^{(\pi)}_{s_d=1} \big]_{j_d=1} 
	\chi_{\frac{1}{2}}(\pi_3) \Big]_S
	\bigg]_{J}^{J_z}
	\big[
	\eta^{(\pi)}_{t_d=0} \; \eta_\frac{1}{2}(\pi_3)\big]_{T=\frac{1}{2}}^{T_z} \,,\label{eq:wf}
\end{align}
with the scattering energy fixed at $E=\frac{q^2}{2\mu_{Nd}}$.
In the expression above, brackets denote the standard coupling of angular momenta with Clebsch-Gordan coefficients: 
the deuteron system, characterized by the quantum numbers $\ell_d=0,2,s_d=1,j_d=1,t_d=0$, is coupled with the incident nucleon to the total spin (isospin) $S$ ($T$); then $S$ is coupled with the relative orbital momentum $L$ to the total momentum $J$. The spinors $\chi_{\frac{1}{2}s_{\pi_1}},\chi_{\frac{1}{2}s_{\pi_2}}$ ($\eta_{\frac{1}{2}t_{\pi_1}},\eta_{\frac{1}{2}t_{\pi_2}}$) relative to the single particles $\pi_1,\pi_2$, respectively, are coupled as follows:
$\chi^{(\pi)}_{s_d,m_{s_d}} \equiv [\chi_{\frac{1}{2}}(\pi_1) \chi_{\frac{1}{2}}(\pi_2)]_{s_d}^{m_{s_d}}$ 
($\eta^{(\pi)}_{t_d,m_{t_d}} \equiv [\eta_{\frac{1}{2}}(\pi_1) \eta_{\frac{1}{2}}(\pi_2)]_{t_d}^{m_{t_d}}$).
Furthermore, we have denoted with $\varphi_{0(2)}$ the deuteron wave function in the $S$ state ($D$ state).
Due to the adopted normalization convention, we have also used the definition $\tilde{\delta} \equiv (2\pi)^3 \delta$.
The wave function~(\ref{eq:wf}) is antisymmetric with respect to the exchange of the particles $\pi_1\pi_2$ constituting the deuteron. Therefore, it is sufficient to consider only the even permutations.

Within this formalism, the matrix elements of the three-body N4LO potential between the $N-d$ scattering states reduce to
\begin{align}
	\langle  \Psi_{L'S'T'}^{JJ_zT_z} | \hat{V} |  \Psi_{LST}^{JJ_zT_z}\rangle 
	= \frac{1}{3} \sum_{\pi=1}^3
	\phantom{\rangle}_{1}\langle  \Psi_{L'S'T'}^{JJ_zT_z} | \hat{V} |  \Psi_{LST}^{JJ_zT_z}\rangle_{\pi} \,.\label{eq:<V>1pi}
\end{align}
By assuming for the three-body interaction the operator structure in Eq.~(\ref{eq:1}), the integrals to be calculated are of the following type:
\begin{align}
	\phantom{\rangle}_{1} \langle \Psi_{L'S'T'}^{JJ_zT_z} | \hat{O}_i | \Psi_{LST}^{JJ_zT_z} \rangle_{\pi} 
	&= 
	\int dp' \delta(p^{\prime}-q') \int d\mathbf{\hat{p}}'
	\sum_{\ell'_d=0,2} \frac{1}{(2\pi)^3} \int dk' k'^{2} \varphi_{\ell'_d}(k') \int d\mathbf{\hat{k}}' \nonumber\\
	&\times\int dp^{(\pi)}  \delta(p^{(\pi)}-q)  \int d\mathbf{\hat{p}}^{(\pi)}
	\sum_{\ell_d=0,2} \frac{1}{(2\pi)^3} \int dk^{(\pi)} (k^{(\pi)})^{2} \varphi_{\ell_d}(k^{(\pi)}) \int d\mathbf{\hat{k}}^{(\pi)} 
	\nonumber\\
	&\times \bigg[ Y_{L'}(\mathbf{\hat{p}}') \Big[  \big[
	Y_{\ell'_d}(\mathbf{\hat{k}}') \; \chi_{1} \big]_{1} 
	\chi_{\frac{1}{2}} \Big]_{S'}
	\bigg]_{J}^{J_z\dagger}
	\big[ \eta_{0} \; \eta_\frac{1}{2} \big]_{T'=\frac{1}{2}}^{T_z \dagger}
	\nonumber\\
	&\times
	O_i^{\left<1\pi\right>}
	\bigg[ Y_L(\mathbf{\hat{p}}^{(\pi)}) \Big[  \big[
	Y_{\ell_d}(\mathbf{\hat{k}}^{(\pi)}) \; \chi^{(\pi)}_{1} \big]_{1} 
	\chi_{\frac{1}{2}}(\pi_3) \Big]_S
	\bigg]_{J}^{J_z}
	\big[ \eta^{(\pi)}_{0} \; \eta_\frac{1}{2}(\pi_3)\big]_{T=\frac{1}{2}}^{T_z}\,. \label{eq:O1pi}
\end{align}
To simplify the notation, superscripts have been omitted when referring to the permutation $\pi=1$, corresponding to the ordering $(123)$.
To give an example, for the operator $O_1$ associated to the LEC $E_1$ [see Eq.~(\ref{eq:1a})] we have explicitly
\begin{subequations}
\begin{align}
	O_1^{\left<1\pi\right>} 
	&= 2[(\mathbf{p}'_1 - \mathbf{p}_{\pi_1})^2 + (\mathbf{p}'_2 - \mathbf{p}_{\pi_2})^2 + (\mathbf{p}'_3 - \mathbf{p}_{\pi_3})^2 ] \,, \\
	&= 3 (p'^2 + p^{(\pi)2}) + 4 (k'^2 + k^{(\pi)2}) - 6 \left( y_\pi - \frac{1}{2} x_\pi \right) \mathbf{p}' \cdot \mathbf{p}^{(\pi)} \,,\label{eq:o1}
\end{align}
\end{subequations}
with the definitions $x_\pi=0,1,-1$ and $y_\pi=1,0,-1$ for $\pi=1,2,3$, respectively.
To obtain the last identity, we have used the relations
\begin{align}
	\mathbf{p}_{\pi_1} &= -\frac{\mathbf{p}^{(\pi)}}{2} - \mathbf{k}^{(\pi)} \,,\\
	\mathbf{p}_{\pi_2} &= -\frac{\mathbf{p}^{(\pi)}}{2} + \mathbf{k}^{(\pi)} \,,\\
	\mathbf{p}_{\pi_3} &= \mathbf{p}^{(\pi)} \,,
\end{align}
which can be easily deduced from Eqs.~(\ref{eq:kp_k}) and~(\ref{eq:kp_p}).
The expressions for the remaining operators can be derived in a similar way~\cite{Riso}.
In any case, the most general operator that we deal with is a
combination of momentum, spin and isospin parts, $O^{m,\sigma} \otimes O^{\tau}$:
the momentum-spin part is a combination of
$\set{ \mathbf{p}'\mathbf{p}',
	\mathbf{p}' \mathbf{p}^{(\pi)},
	\mathbf{p}^{(\pi)} \mathbf{p}^{(\pi)}, 
	\mathbf{k}'\mathbf{k}', 
	\mathbf{k}^{(\pi)}\mathbf{k}^{(\pi)}}$
and
$\set{\operatorname{I}^\sigma,\bmsi_i}_{i=1,2,3}$,
taken as scalar or tensorial structures (e.g.: $O^{m,\sigma} \propto p'^2$, 
$p'^2 (\sigma_1 \sigma_2)_0$,
$[(\mathbf{p}' \mathbf{p}^{(\pi)})_1 \sigma_1]_0$,
$[(\mathbf{p}'\mathbf{p}')_2 (\sigma_1 \sigma_2)_2]_0$, $\dots$); 
the isospin part is given by $O^\tau = \set{\operatorname{I}^\tau,\bmta_i \cdot \bmta_j}_{ij=12,23,31}$.

In the most general case, the action of a given operator on the state
\begin{align}
		|J,\ell_d,L,S,T;J_z,T_z\rangle_\pi \equiv
		\left|
		\bigg[ Y_L(\mathbf{\hat{p}}^{(\pi)}) \Big[  \big[
		Y_{\ell_d}(\mathbf{\hat{k}}^{(\pi)}) \; \chi^{(\pi)}_{1} \big]_{1} 
		\chi_{\frac{1}{2}}(\pi_3) \Big]_S
		\bigg]_{J}^{J_z}
		\big[
		\eta^{(\pi)}_{0} \; \eta_\frac{1}{2}(\pi_3)\big]_{\frac{1}{2}}^{T_z} 
		\right\rangle_{\pi} \,,
\end{align}
results in a linear combination of states  having the same $J,J_z$ ($T_z$), for angular-momentum (isospin) conservation, and the parity conservation implies $(-)^{L+L'}=1$. Moreover, through a recombination of quantum numbers, it is always possible to express these states in terms of a  ``reference'' ordering, here $\pi=1$.
Hence, we need to evaluate 
\begin{equation}
		\hat{Q} |J,\ell_d,L,S,T;J_z,T_z\rangle_\pi
		= \sum_{S'} \alpha^{Q}_{\pi}(\ell_d,L,S,\ell'_d=0,L'=0,S')  \;
		|J,\ell''_d,L'',S',T;J_z,T_z\rangle_1  + \dots \label{eq:alpha}
\end{equation}
with $\hat{Q}$ an  ``elementary'' operator appearing in each $O_1,\dots,O_{13}$ of Eq.~(\ref{eq:1}).
Note that the integrals over the variables $\mathbf{\hat{p}}^{(\pi)}$ and $\mathbf{\hat{k}}^{(\pi)}$ are implied.
Furthermore, states in which, for example, the spins of the first and the second particles combine to give $s_d=0$ are excluded from the summation and accounted for by the dots. This is because these states do not contribute to matrix elements with a bra in the ordering $\pi=1$, where $s_d=1$ states are present.
The coefficients $\alpha^{Q}_{\pi}$, whose detailed calculation is given in Appendix~\ref{secA1}, are used to fully determine the potential matrix elements~(\ref{eq:O1pi}).
Since these matrix elements also depend on factors that are functions of the deuteron structure only, it is useful to define the following integrals:
\begin{eqnarray}
	D_0 &=&\frac{1}{2\pi^2} \int d k k^2 \varphi_0(k) \,,\label{D0}\\
	D_0^{\prime} &=&\frac{1}{2\pi^2} \int d k k^4 \varphi_0(k) \,,\label{D0prime}\\
	D_2 &=&\frac{1}{2\pi^2} \int d k k^4 \varphi_2(k)\,. \label{D2}
\end{eqnarray} 

In the rest of this section, we present the results relative to the calculation of the potential matrix elements for the scattering channels $J^\pi=\frac{1}{2}^\pm,\frac{3}{2}^\pm,\frac{5}{2}^\pm$.
The spectroscopic notation ${}^{2S+1}L_J$ is employed and only the matrix elements that are different from zero are listed.
For each matrix element, we also define the so-called channel constants, $E^{(c)}_i$, which are linearly related to combinations of the potential LECs $E_1,\dots,E_{13}$. 

$J^\pi=\frac{1}{2}^+$:
\begin{subequations}
	\begin{align}
		\left\langle{ }^{2} S_{1/2} \middle|\hat{V}\middle|^{2} S_{1/2}\right\rangle
		&=-E_1 \bigg[ 4 D_0 D'_0 + \frac{3}{2} D_0^2(q'^2+q^2) \bigg]
		+E_2 \bigg[ 6 D_0 D'_0 + \frac{3}{4} D_0^2(q'^2+q^2) \bigg] \nonumber\\
		&+E_3 \bigg[ 2 D_0 D'_0 + \frac{9}{4} D_0^2(q'^2+q^2) \bigg]
		+E_4 \bigg[12 D_0 D'_0 + \frac{9}{2} D_0^2(q'^2+q^2) \bigg] \nonumber\\
		&-2\sqrt{2} E_5 D_0 D_2  +6\sqrt{2} E_6 D_0 D_2  \nonumber\\
		&+E_9 \bigg[ \frac{2}{3} D_0 D'_0 +\frac{4\sqrt{2}}{3} D_0 D_2 - \frac{3}{4} D_0^2(q'^2+q^2)\bigg]  \nonumber\\
		&-E_{10} \bigg[ 2 D_0 D'_0 + 4\sqrt{2} D_0 D_2 + \frac{3}{4} D_0^2(q'^2+q^2)\bigg] \nonumber\\
		&+E_{11} \bigg[ \frac{2}{3} D_0 D'_0 +\frac{4\sqrt{2}}{3} D_0 D_2 - \frac{3}{4} D_0^2(q'^2+q^2)\bigg]  \nonumber\\
		&-E_{12} \bigg[ 2 D_0 D'_0 + 4\sqrt{2} D_0 D_2 + \frac{3}{4} D_0^2(q'^2+q^2)\bigg] \nonumber\\
		&+ \frac{3}{2}   E_{13} D_0^2 (q'^2+q^2)
		\\
		&\equiv \widetilde{E}^{(c)}_{{ }^{2} S_{1/2}} + E^{(c)}_{{ }^{2} S_{1/2}} (q'^2+q^2)
	\end{align}
\end{subequations}
\begin{subequations}
\begin{align}
	\left\langle{ }^{2} S_{1/2} \middle|\hat{V}\middle|^{4} D_{1/2}\right\rangle
	&=\frac{1}{\sqrt{2}} \left( \frac{3}{2}E_5 - \frac{9}{2}E_6
	- E_9 + 3 E_{10} - E_{11} + 3E_{12}\right) D_0^2 q^2 
	\\
	&\equiv E^{(c)}_{1/2^+} q^2
\end{align}
\end{subequations}

$J^\pi=\frac{1}{2}^-$:
\begin{subequations}
\begin{align}
	\left\langle{}^{2} P_{1/2} \middle|\hat{V}\middle|{}^{2} P_{1/2}\right\rangle
	&=\frac{1}{2}\bigg( E_1 - E_2 - E_3 + 3 E_4
	+ \frac{1}{2}E_7 -\frac{1}{2}E_8  \nonumber\\
	&+ \frac{7}{6}E_9 - \frac{5}{2}E_{10}
	- \frac{5}{6}E_{11} - \frac{1}{2}E_{12} 
	+ \frac{3}{2}E_{13} \bigg) D_0^2 q'q
	\\
	&\equiv E^{(c)}_{{ }^{2} P_{1/2}} q'q
\end{align}
\end{subequations}
\begin{subequations}
\begin{align}
	\left\langle{}^{4} P_{1/2} \middle|\hat{V}\middle|{}^{4} P_{1/2}\right\rangle
	&=\bigg( E_1 + E_3 -5 E_5
	+ \frac{5}{4}E_7-\frac{15}{4}E_8  \nonumber\\
	&+\frac{2}{3}E_9 +2E_{10}+\frac{2}{3}E_{11}+2E_{12} -2E_{13}
	\bigg) D_0^2 q'q
	\\
	&\equiv E^{(c)}_{{}^4P_{1/2}}q'q
\end{align}
\end{subequations}
\begin{subequations}
\begin{align}
	\left\langle{}^{2} P_{1/2} \middle|\hat{V}\middle|{}^{4} P_{1/2}\right\rangle
	&=\frac{1}{\sqrt{2}} \bigg( -\frac{5}{2}E_5 - \frac{15}{2}E_6
	+ \frac{1}{4}E_7 - \frac{1}{2}E_8  \nonumber\\
	&+ \frac{5}{3}E_9 - E_{10}
	+ \frac{5}{3}E_{11} + E_{12} - 3E_{13}
	\bigg) D_0^2 q'q
	\\
	&\equiv E^{(c)}_{{1/2}^-}q'q
\end{align}
\end{subequations}

$J^\pi=\frac{3}{2}^-$:
\begin{subequations}
\begin{align}
	\left\langle{}^{2} P_{3/2} \middle|\hat{V}\middle|{}^{2} P_{3/2}\right\rangle
	&=\frac{1}{2} \bigg( E_1- E_2 - E_3 + 3 E_4
	- \frac{1}{4}E_7 + \frac{1}{4} E_8  \nonumber\\
	&- \frac{1}{3} E_9 - E_{10} + \frac{2}{3} E_{11} - 2 E_{12}
	\bigg) D_0^2 q'q
	\\
	&\equiv E^{(c)}_{{}^2P_{3/2}} q'q
\end{align}
\end{subequations}
\begin{subequations}
\begin{align}
	\left\langle{}^{4} P_{3/2} \middle|\hat{V}\middle|{}^{4} P_{3/2}\right\rangle
	&=\bigg( E_1+ E_3 +4 E_5
	+ \frac{1}{2} E_7 - \frac{3}{2} E_8  \nonumber\\
	&- \frac{5}{6}E_9 - \frac{5}{2}E_{10}
	- \frac{5}{6}E_{11} - \frac{5}{2} E_{12} + \frac{5}{2} E_{13}
	\bigg) D_0^2 q'q
	\\
	&\equiv E^{(c)}_{{}^4P_{3/2}}q'q
\end{align}
\end{subequations}
\begin{subequations}
\begin{align}
	\left\langle{}^{2} P_{3/2} \middle|\hat{V}\middle|{}^{4} P_{3/2}\right\rangle
	&=\frac{\sqrt{5}}{2} \bigg( \frac{1}{2} E_5 + \frac{3}{2} E_6 + \frac{1}{4}E_7 -\frac{1}{2} E_8
	- \frac{1}{3} E_9
	-E_{10}
	- \frac{1}{3} E_{11} +  E_{12}
	\bigg) D_0^2 q'q
	\\
	&\equiv E^{(c)}_{{3/2}^-}q'q
\end{align}
\end{subequations}

$J^\pi=\frac{5}{2}^-$:
\begin{subequations}
\begin{align}
	\left\langle{}^{4} P_{5/2} \middle|\hat{V}\middle|{}^{4} P_{5/2}\right\rangle
	&=\bigg( E_1 + E_3 - E_5 -\frac{3}{4} E_7 + \frac{9}{4} E_8
	\bigg) D_0^2 q'q
	\\
	&\equiv E^{(c)}_{{}^4P_{5/2}}q'q
\end{align}
\end{subequations}
In summary, 10 constants, $E^{(c)}_i$, have been defined: three for each of the scattering channels $J^\pi=\frac{1}{2}^+,\frac{1}{2}^-,\frac{3}{2}^-$, and one for the channel $\frac{5}{2}^-$.
An additional combination of LECs $E_1,\dots,E_{13}$ should emerge when taking into account momentum operator structures of the type $\set{\mathbf{p}'\mathbf{k}', \mathbf{p}'\mathbf{k}^{(\pi)},\mathbf{k}'\mathbf{p}^{(\pi)},\mathbf{k}^{(\pi)}\mathbf{p}^{(\pi)},\mathbf{k}'\mathbf{k}^{(\pi)}}$. These have been excluded here, since they do not contribute to the presented calculation, for which the quantum numbers of the particle pair $\pi_1\pi_2$ have been fixed to those of the deuteron.

\section{Conclusions}\label{sec:conclusions}
We carried out an analysis of the most general form of subleading three-nucleon contact interactions at N4LO, focusing on their isospin projection and partial-wave decomposition.

Regarding the isospin-projection study, we explicitly constructed 13 combinations of operators and classified them according to their isospin structure, separating the pure $T = \frac{1}{2}$ from the $T = \frac{3}{2}$ components.
We found that 11 constants, $E^{(1/2)}_{i}$, are relevant for the total isospin $T=\frac{1}{2}$ sector, and that two constants, $E^{(3/2)}_{i}$, contribute to the $T=\frac{3}{2}$ sector.
These results confirm that 11 linearly independent combinations of LECs $E_1,\dots,E_{13}$ contribute to $N-d$ observables, in agreement with symmetry arguments. This is consistent with Ref.~\cite{PhysRevC.105.054004_Witala}.

We then performed a partial-wave decomposition of the potential terms in the $N-d$ elastic scattering channels. 
When calculating the matrix elements, 10 different combinations of $E_1,\dots,E_{13}$ were identified and denoted as $E^{(c)}_i$. 
For each of the channels $J^\pi=\frac{1}{2}^+,\frac{1}{2}^-,\frac{3}{2}^-$ three constants were found, while only one for the $\frac{5}{2}^-$ state.
The absence of an additional combination of $E_1,\dots,E_{13}$ should be attributed to the fact that not all the momentum operator structures were taken into account in the calculation, since some of them give no contribution when deuteron quantum numbers are imposed on the assumed $N-d$ scattering wave function. The 10 channel constants would give the bulk of the subleading contact contribution provided they can be treated perturbatively. Using more general scattering wave functions, as in Ref.~\cite{zuo2025threenucleoncontactforcesjacobi}, all of the 11 $E^{(1/2)}_i$ should contribute.

The presented partial-wave decomposition of the subleading 3N contact interaction provides a starting point to derive explicit expressions for $p-d$ scattering observables provided their effect can be treated perturbatively.
In order to carry out a proper phase-shift analysis and study the impact of the N4LO three-body contact potential, a different assumption should be made concerning the $N-d$ scattering wave function $\Psi_{LST}^{JJ_zT_z}$. As the scattering channels involved have a relative orbital momentum of $L\le2$ between the deuteron and the incident nucleon, the complete wave function should be considered, with the core part no longer being neglected.
In this analysis, we expect an overall dependence of the results on the 11 expected channel constants, since all the momentum operator structures should be considered when the core wave function is included in the calculation.

This analysis provides a tool to identify the combinations of LECs more sensitive to specific channels of elastic $p-d$ scattering and to guide the fit procedures to experimental observables \cite{Girlanda:2018xrw,Filandri2023,PhysRevC.105.054004_Witala} making use of existing phaseshift analyses \cite{Knutson:1993zz,Kievsky:1996ca,Wood:2001gb}.

\appendix
\section{}\label{secA1}
In this appendix we derive the coefficients $\alpha^Q_\pi$ as defined in Eq.~(\ref{eq:alpha}).
For simplicity, we will omit the dots from the equations defining the different $\alpha^Q_\pi$, as well as the quantum numbers $J_z,T_z$ in denoting the states.

Since the isospin part always factorizes, it is convenient to define coefficients related only to the isospin operators $\operatorname{I^\tau}, \bmta_1 \cdot \bmta_2, \bmta_2 \cdot \bmta_3, \bmta_1 \cdot \bmta_3$. We introduce a matrix of coefficients $\alpha_{\pi, j}^{\tau}$, where the first lower index refers to the permutation and the second index, $j=0, \ldots, 3$, refers to one of the aforementioned four operators, so that
\begin{align}
	\operatorname{I^\tau}|T=1/2 \rangle_\pi
	&=\alpha_{\pi, 0}^{\tau}|T=1/2\rangle_{1} \,,\\
	\bmta_1 \cdot \bmta_2|T=1/2\rangle_\pi 
	&=\alpha_{\pi, 1}^{\tau}|T=1/2\rangle_{1} \,,\\
	\bmta_2 \cdot \bmta_3|T=1/2\rangle_\pi 
	&=\alpha_{\pi, 2}^{\tau}|T=1/2\rangle_{1} \,,\\
	\bmta_1 \cdot \bmta_3|T=1/2\rangle_\pi 
	&=\alpha_{\pi, 3}^{\tau}|T=1/2\rangle_{1} \,.
\end{align}
Hence, we have
\begin{align}
	&\alpha_{1,0}^{\tau}=1\,, &&\alpha_{1,1}^{\tau}=-3\,, &&\alpha_{1,2}^{\tau}=0\,, &&\alpha_{1,3}^{\tau}=0 \,, \\
	&\alpha_{2,0}^{\tau}=-\frac{1}{2}\,, &&\alpha_{2,1}^{\tau}=\frac{3}{2}\,, &&\alpha_{2,2}^{\tau}=\frac{3}{2}\,, &&\alpha_{2,3}^{\tau}=-\frac{3}{2} \,,\\
	&\alpha_{3,0}^{\tau}=-\frac{1}{2}\,, &&\alpha_{3,1}^{\tau}=\frac{3}{2}\,,  &&\alpha_{3,2}^{\tau}=-\frac{3}{2}\,, &&\alpha_{3,3}^{\tau}=\frac{3}{2} \,.
\end{align}

The simplest operator is the spin identity $\operatorname{I^\sigma}$. 
Defining the coefficients $\alpha_{\pi,0}^{\sigma}$ such that
\begin{equation}
	\operatorname{I^\sigma} |J, \ell_d=0, L=0, S\rangle_\pi = (4\pi)^2 \alpha_{\pi,0}^{\sigma}  |J, \ell_d=0, L=0, S\rangle_{1} \,,
\end{equation}
one has:
\begin{align}
	\alpha_{1,0}^{\sigma} &=1 \,, \\
	\alpha_{2,0}^{\sigma} &=(-1)^{\frac{3}{2}-S} T_{1,1,S}^{\frac{1}{2}, \frac{1}{2}, \frac{1}{2}} \,, \\
	\alpha_{3,0}^{\sigma}&=(-1)^{\frac{3}{2}-S} T_{1,1,S}^{\frac{1}{2}, \frac{1}{2}, \frac{1}{2}} \,.
\end{align}
The $T$ symbols refer to the recoupling of three angular momenta through Wigner $6j$ coefficients~\cite{Edmonds}, 
\begin{equation}
	\big[(j_1 j_2)_{j_{12}} j_3\big]_j 
	= \sum_{j_{23}} T^{j_1,j_2,j_3}_{j_{12}, j_{23}, j} \big[ j_1(j_2 j_3)_{j_{23}}\big]_j  \,,
\end{equation}
and are defined as
\begin{equation}
	T^{j_1,j_2,j_3}_{j_{12}, j_{23}, j} = (-1)^{j_1+j_2+j_3+j} \widehat{j_{12}} \widehat{j_{23}} \begin{Bmatrix}j_1 & j_2 & j_{12}\\j_3 & j & j_{23}\end{Bmatrix} \,,
\end{equation}
with $\widehat{x} \equiv \sqrt{2x+1}$.
Similarly, we define the action of the spin operators $\bmsi_i \cdot \bmsi_j \equiv -\sqrt{3} \left( \sigma_i \sigma_j \right)_0$  in terms of the coefficients $\alpha_{\pi, j}^\sigma$, with $j=1,2,3$. More precisely:
\begin{align}
	\left(\sigma_1 \sigma_2\right)_0 \big|J, \ell_d=0,L=0, S \big\rangle_\pi
	&=(4\pi)^2 \alpha_{\pi, 1}^\sigma \big|J,\ell_d=0, L=0, S \big\rangle_{1} \,,\\
	\left(\sigma_2 \sigma_3\right)_0 \big|J,\ell_d=0, L=0, S \big\rangle_\pi
	&=(4\pi)^2 \alpha_{\pi, 2}^\sigma \big|J, \ell_d=0,L=0, S \big\rangle_{1} \,,\\
	\left(\sigma_1 \sigma_3\right)_0 \big|J, \ell_d=0,L=0, S \big\rangle_\pi
	&=(4\pi)^2 \alpha_{\pi, 3}^\sigma \big|J, \ell_d=0,L=0, S \big\rangle_{1}  \,,
\end{align}
with
\begin{align}
	&\alpha_{1,1}^\sigma=3 N_{0,1, \frac{1}{2}, \frac{1}{2}, 1}^{1,1, \frac{1}{2}, \frac{1}{2}}=1\,,\\
	&\alpha_{1,2}^\sigma=3 N_{0, S, 1, \frac{1}{2}, S}^{1,1,1, \frac{1}{2}} T_{\frac{1}{2}, 1,1}^{1, \frac{1}{2}, \frac{1}{2}}\,,\\
	&\alpha_{1,3}^\sigma=\alpha_{1,2}^\sigma\,,\\
	&\alpha_{2,1}^\sigma=3(-1)^{\frac{3}{2}-S} T_{1,1, S}^{\frac{1}{2}, \frac{1}{2}, \frac{1}{2}} N_{0,1, \frac{1}{2}, \frac{1}{2}, 1}^{1,1, \frac{1}{2},\frac{1}{2}} \,,\\
	&\alpha_{2,2}^\sigma=\alpha_{2,1}^\sigma\,,\\
	&\alpha_{2,3}^\sigma=3 \sum_{\tilde{s}=0}^1(-1)^{\frac{3}{2}-S} N_{0, S, \tilde{s}, \frac{1}{2}, S}^{1,1,1, \frac{1}{2}} T_{\frac{1}{2}, 1, \tilde{s}}^{1, \frac{1}{2}, \frac{1}{2}} T_{\tilde{s}, 1, S}^{\frac{1}{2}, \frac{1}{2}, \frac{1}{2}}\,,\\
	&\alpha_{3,1}^\sigma=3 \sum_{\tilde{s}=0}^1(-1)^{\tilde{s}+\frac{1}{2}-S} N_{0, S, \tilde{s}, \frac{1}{2}, S}^{1,1,1, \frac{1}{2}} T_{\frac{1}{2}, 1, \tilde{s}}^{1, \frac{1}{2}, \frac{1}{2}} T_{1, \tilde{s}, S}^{\frac{1}{2}, \frac{1}{2}, \frac{1}{2}} \text {, }\\
	&\alpha_{3,2}^\sigma=\alpha_{2,3}^\sigma \,,\\
	&\alpha_{3,3}^\sigma=\alpha_{2,1}^\sigma \,.
\end{align}
The $N$ symbols introduced above are related to the recoupling of four angular momenta, 
\begin{equation}
	\big[(j_1 j_2)_{j_{12}} (j_3 j_4)_{j_{34}}\big]_j 
	= \sum_{j_{13} j_{24}} N^{j_1,j_2,j_3,j_4}_{j_{12}, j_{34}, j_{13}, j_{24},j} \big[(j_1 j_3)_{j_{13}} (j_2 j_4)_{j_{24}}\big]_j  \,,
\end{equation}
and they are given in terms of Wigner $9j$ coefficients~\cite{Edmonds} as follows:
\begin{equation}
	N^{j_1,j_2,j_3,j_4}_{j_{12}, j_{34}, j_{13}, j_{24}, j} 
	= \widehat{j_{12}} \widehat{j_{34}} \widehat{j_{13}} \widehat{j_{24}} \begin{Bmatrix}j_1 & j_2 & j_{12}\\j_3 & j_4 & j_{34}\\j_{13} & j_{24} & j\end{Bmatrix} \,.
\end{equation}
For example, making use of the above definitions, the potential matrix element (\ref{eq:<V>1pi}) corresponding to the operator $O_1$ [see Eq.(\ref{eq:o1})] can be expressed as follows:
\begin{align}
	\langle \Psi_{L'S'\frac{1}{2}}^{J,J_z,T_z} | \hat{O}_1 | \Psi_{LS\frac{1}{2}}^{J,J_z,T_z} \rangle 
	&= \delta_{S'S} \bigg\{ \delta_{L'0} \delta_{L0}  \bigg[ \frac{8}{3} D'_0 D_0 + D_0^2(q'^2+q^2) \bigg] \sum_{\pi=1}^3 \alpha_{\pi,0}^{\sigma} \alpha_{\pi,0}^{\tau} \nonumber\\
	&+ \delta_{L'1} \delta_{L1}
	\frac{2}{\sqrt{3}} T^{1,1,1}_{0,0,1} B^0_{1,1} D_0^2 q'q \sum_{\pi=1}^3 \bigg( y_\pi - \frac{1}{2} x_\pi \bigg) \alpha_{\pi,0}^{\sigma} \alpha_{\pi,0}^{\tau} \bigg\} \,,
\end{align}
with $D_0,D'_0$ as defined in Eqs.~(\ref{D0}) and~(\ref{D0prime}).
$B$ symbols denote the coupling of two spherical harmonics with the same argument,
\begin{equation}
	[Y_{\ell_1},Y_{\ell_2}]_\lambda^m = \frac{B^{\lambda}_{\ell_1,\ell_2}}{\sqrt{4\pi}} Y_\lambda^m \,,
\end{equation}
and they depend on the Wigner 3j symbols as follows~\cite{Edmonds}:
\begin{equation}
	B^{\lambda}_{\ell_1,\ell_2} = (-1)^{\ell_1+\ell_2} \hat{\ell_1} \hat{\ell_2} \begin{pmatrix}\ell_1&\ell_2&\lambda\\0&0&0\end{pmatrix} \,.
\end{equation}
The matrix elements of the operators $O_2,O_3,O_4$ can be calculated similarly.

The action of the tensor operators that combine spin variables and orbital variables, of the kind of $O_5, \dots,O_{13}$, depends on whether they act on $S$, $P$ or $D$ waves. 
In the first case these are
\begin{align}
	&\big[(\mathbf{\hat{k}} \mathbf{\hat{k}})_2\left(\sigma_1 \sigma_2\right)_2\big]_0 \big| J, \ell_d=2, L=0, S\big\rangle_\pi
	=(4\pi)^2 \alpha_{\pi, 1}^{T 0} \big|J, \ell_d=0, L=0, S\big\rangle_{1} \,,
	\\
	&\big[(\mathbf{\hat{k}} \mathbf{\hat{k}})_2 \left(\sigma_2 \sigma_3\right)_2\big]_0 \big|J, \ell_d=2, L=0, S \big\rangle_\pi
	=(4\pi)^2 \alpha_{\pi, 2}^{T 0} \big|J, \ell_d=0, L=0, S \big\rangle_{1} \,,
	\\
	&\big[(\mathbf{\hat{k}} \mathbf{\hat{k}})_2\left(\sigma_1 \sigma_3\right)_2\big]_0 \big|J, \ell_d=2, L=0, S \big\rangle_\pi
	=(4\pi)^2 \alpha_{\pi, 3}^{T 0} \big|J, \ell_d=0, L=0, S \big\rangle_{1} \,,
\end{align}
with
\begin{align}
	\alpha_{1,1}^{T 0}
	&=B_{1,1}^2 B_{2,2}^0 N_{0,1,0,1,1}^{2,2,2,1} N_{2,1,\frac{1}{2}, \frac{1}{2},1}^{1,1,\frac{1}{2},\frac{1}{2}}\,,\\
	\alpha_{1,2}^{T 0}
	&=B_{1,1}^2 B_{2,2}^0 N_{0, S, 1, \frac{1}{2}, S}^{1,1,1, \frac{1}{2}} N_{1,1,0,1,1}^{2,1,2,1} T_{\frac{1}{2}, 1,1}^{1, \frac{1}{2}, \frac{1}{2}}\,,\\
	\alpha_{1,3}^{T 0}
	&=\alpha_{1,2}^{T 0}\,,\\
	\alpha_{2,1}^{T 0}
	&=B_{1,1}^2 B_{2,2}^0 \sum_{\tilde{s}=0}^1(-1)^{\frac{1}{2}+\tilde{s}-S} N_{0, S, \tilde{s}, \frac{1}{2}, S}^{1,1,1, \frac{1}{2}}  N_{1,1,0, \tilde{s}, S}^{2,1,2,1} T_{\frac{1}{2}, 1, \tilde{s}}^{1, \frac{1}{2}, \frac{1}{2}} T_{1, \tilde{s}, S}^{\frac{1}{2}, \frac{1}{2}, \frac{1}{2}}\,,\\
	\alpha_{2,2}^{T 0}
	&=B_{1,1}^2 B_{2,2}^0 N_{0,1,0,1,1}^{2,2,2,1} N_{2,1, \frac{1}{2}, \frac{1}{2}, 1}^{1,1, \frac{1}{2},\frac{1}{2}} T_{1,1,S}^{\frac{1}{2}, \frac{1}{2}, \frac{1}{2}}\,,\\
	\alpha_{2,3}^{T 0}
	&=B_{1,1}^2 B_{2,2}^0 \sum_{\tilde{s}=0}^1(-1)^{\frac{3}{2}-S} N_{0, S, \tilde{s}, \frac{1}{2}, S}^{1,1,1, \frac{1}{2}} N_{1,1,0, \tilde{s}, S}^{2,1,2,1}
	T_{\frac{1}{2}, 1, \tilde{s}}^{1, \frac{1}{2}, \frac{1}{2}} T_{ \tilde{s},1, S}^{\frac{1}{2}, \frac{1}{2}, \frac{1}{2}}\,,\\
	\alpha_{3,1}^{T 0}
	&=\alpha_{2,1}^{T 0}\,,\\
	\alpha_{3,2}^{T 0}
	&=\alpha_{2,3}^{T 0}\,,\\
	\alpha_{3,3}^{T 0}
	&=(-1)^{\frac{3}{2}-S} B_{1,1}^2 B_{2,2}^0 N_{0,1,0,1,1}^{2,2,2,1} N_{2,1, \frac{1}{2}, \frac{1}{2}, 1}^{1,1, \frac{1}{2},\frac{1}{2}} T_{1,1, S}^{\frac{1}{2}, \frac{1}{2}, \frac{1}{2}} \,.
\end{align}
The same operators with $\mathbf{\hat{k}}$ replaced by $\mathbf{\hat{k}}'$ will contribute to the matrix elements $\langle \ell_d=2 | \hat{V} | \ell_d=0 \rangle$.

The tensor operators acting on the $P$ waves give
\begin{align}
	\big[\left(\mathbf{\hat{p}}^{\prime} \mathbf{\hat{p}}\right)_2\left(\sigma_1 \sigma_2\right)_2\big]_0 \big|J, \ell_d=0, L=1, S\big\rangle_\pi
	&=(4\pi)^2 \sum_{S^{\prime}} \alpha_{\pi, 1}^{T 1}\left(S, S^{\prime}\right)
	\big|J, \ell_d=0, L=1, S^{\prime}\big\rangle_{1} \,,
	\\
	\big[\left(\mathbf{\hat{p}}^{\prime} \mathbf{\hat{p}}\right)_2\left(\sigma_2 \sigma_3\right)_2\big]_0 \big|J, \ell_d=0, L=1, S\big\rangle_\pi
	&=(4\pi)^2 \sum_{S^{\prime}} \alpha_{\pi, 2}^{T 1}\left(S, S^{\prime}\right)
	\big|J, \ell_d=0, L=1, S^{\prime}\big\rangle_{1} \,,
	\\
	\big[\left(\mathbf{\hat{p}}^{\prime} \mathbf{\hat{p}}\right)_2\left(\sigma_1 \sigma_3\right)_2\big]_0 \big|J, \ell_d=0, L=1, S\big\rangle_\pi 
	&=(4\pi)^2 \sum_{S^{\prime}} \alpha_{\pi, 3}^{T 1}\left(S, S^{\prime}\right)
	\big|J, \ell_d=0, L=1, S^{\prime}\big\rangle_{1} \,,
\end{align}
where
\begin{align}
	\alpha_{1,1}^{T 1}\left(S, S^{\prime}\right)
	&=B_{1,1}^0 T_{2,0,1}^{1,1,1} N_{0, J, 1, S', J}^{2,2,1,S} T_{1, S, S'}^{2,1, \frac{1}{2}} N_{2,1, \frac{1}{2}, \frac{1}{2}, 1}^{1,1, \frac{1}{2},\frac{1}{2}} \,,
	\\
	\alpha_{1,2}^{T 1}\left(S, S^{\prime}\right)
	&=B_{1,1}^0 T_{2,0,1}^{1,1,1} N_{0, J, 1, S^{\prime}, J}^{2,2,1, S} N_{2, S, 1, \frac{1}{2}, S^{\prime}}^{1,1,1 \frac{1}{2}} T_{\frac{1}{2}, 1,1}^{1, \frac{1}{2}, \frac{1}{2}} \,,
	\\
	\alpha_{1,3}^{T 1}\left(S, S^{\prime}\right)
	&=\alpha_{1,2}^{T 1}\left(S, S^{\prime}\right) \,,
	\\
	\alpha_{2,1}^{T 1}\left(S, S^{\prime}\right)
	&=B_{1,1}^0 T_{2,0,1}^{1,1,1} N_{0, J, 1, S^{\prime}, J}^{2,2,1, S} \sum_{\tilde{s}=0}^1(-1)^{\frac{1}{2}+\tilde{s}-S^{\prime}}
	N_{2, S, \tilde{s}, \frac{1}{2}, S^{\prime}}^{1,1,1, \frac{1}{2}} T_{\frac{1}{2}, 1, \tilde{s}}^{1, \frac{1}{2},\frac{1}{2}} T_{1, \tilde{s}, S^{\prime}}^{\frac{1}{2}, \frac{1}{2},\frac{1}{2}} \,,
	\\
	\alpha_{2,2}^{T 1}\left(S, S^{\prime}\right)
	&=B_{1,1}^0 T_{2,0,1}^{1,1,1} N_{0, J, 1, S^{\prime}, J}^{2,2,1, S} \sum_{\tilde{s}=0}^1(-1)^{\frac{1}{2}+\tilde{s}-S^{\prime}}
	N_{2,1, \frac{1}{2}, \frac{1}{2}, \tilde{s}}^{1,1, \frac{1}{2},\frac{1}{2}} T_{\tilde{s}, S, S^{\prime}}^{2,1, \frac{1}{2}} T_{1,\tilde{s}, S^{\prime}}^{\frac{1}{2}, \frac{1}{2}, \frac{1}{2}} \,,
	\\
	\alpha_{2,3}^{T1}\left(S, S^{\prime}\right)
	&=B_{1,1}^0 T_{2,0,1}^{1,1,1} N_{0, J, 1, S^{\prime}, J}^{2,2,1, S} \sum_{\tilde{s}=0}^1(-1)^{\frac{3}{2}-S^{\prime}}
	N_{2, S,  \tilde{s}, \frac{1}{2}, S^{\prime}}^{1,1,1, \frac{1}{2}} T_{\frac{1}{2}, 1, \tilde{s}}^{1, \frac{1}{2}, \frac{1}{2}} T_{\tilde{s}, 1, S^{\prime}}^{\frac{1}{2}, \frac{1}{2}, \frac{1}{2}} \,,
	\\
	\alpha_{3,1}^{T 1}\left(S, S^{\prime}\right)
	&=\alpha_{2,1}^{T 1}\left(S, S^{\prime}\right) \,,
	\\
	\alpha_{3,2}^{T 1}\left(S, S^{\prime}\right)
	&=\alpha_{2,3}^{T 1}\left(S, S^{\prime}\right) \,,
	\\
	\alpha_{3,3}^{T 1}\left(S, S^{\prime}\right)
	&=B_{1,1}^0 T_{2,0,1}^{1,1,1} N_{0, J, 1, S^{\prime}, J}^{2,2,1, S} \sum_{\tilde{s}=0}^1(-1)^{\frac{3}{2}-S^{\prime}}
	N_{2,1, \frac{1}{2}, \frac{1}{2}, \tilde{s}}^{1,1, \frac{1}{2},\frac{1}{2}}, T_{\tilde{s}, S, S^{\prime}}^{2,1, \frac{1}{2}} T_{\tilde{s},1, S^{\prime}}^{\frac{1}{2}, \frac{1}{2}, \frac{1}{2}} \,.
\end{align}
As can be seen, these operators can change the total spin.

Tensor operators acting on $D$ waves contributing to mixing with $S$ waves are
\begin{align}
	\big[(\mathbf{\hat{p}} \mathbf{\hat{p}})_2\left(\sigma_1 \sigma_2\right)_2\big]_0 \big|J, \ell_d=0, L=2, S\big\rangle_\pi
	&=(4\pi)^2 \alpha_{\pi, 1}^{T 2} \big|J, \ell_d=0, L=0, S \big\rangle_{1} \,,
	\\
	\big[(\mathbf{\hat{p}} \mathbf{\hat{p}})_2\left(\sigma_2 \sigma_3\right)_2\big]_0 \big|J, \ell_d=0, L=2, S\big\rangle_\pi
	&=(4\pi)^2 \alpha_{\pi, 2}^{T 2} \big|J, \ell_d=0, L=0, S \big\rangle_{1} \,,
	\\
	\big[(\mathbf{\hat{p}} \mathbf{\hat{p}})_2\left(\sigma_1 \sigma_3\right)_2\big]_0 \big|J, \ell_d=0, L=2, S\big\rangle_\pi
	&=(4\pi)^2 \alpha_{\pi, 3}^{T 2} \big|J, \ell_d=0, L=0, S \big\rangle_{1} \,.
\end{align}
Here we have
\begin{align}
	\alpha_{1,1}^{T 2}
	&=B_{1,1}^2 B_{2,2}^0 N_{0, J, 0, J, J}^{2,2,2, S} T_{1, S, J}^{2,1, \frac{1}{2}} N_{2,1, \frac{1}{2}, \frac{1}{2}, 1}^{1,1, \frac{1}{2},\frac{1}{2}},\\
	\alpha_{1,2}^{T 2}
	&=B_{1,1}^2 B_{2,2}^0 N_{0, J, 0, J, J}^{2,2,2, S} N_{2, S, 1, \frac{1}{2}, J}^{1,1,1, \frac{1}{2}} T_{\frac{1}{2}, 1,1}^{1, \frac{1}{2}, \frac{1}{2}},\\
	\alpha_{1,3}^{T 2}
	&=\alpha_{1,2}^{T 2},\\
	\alpha_{2,1}^{T 2} 
	&=B_{1,1}^2 B_{2,2}^0 N_{0, J, 0, J, J}^{2,2,2, S} \sum_{\tilde{s}=0}^1(-1)^{\frac{1}{2}+\tilde{s}-J} N_{2, S, \tilde{s}, \frac{1}{2}, J}^{1,1,1, \frac{1}{2}}
	T_{\frac{1}{2}, 1,  \tilde{s}}^{1, \frac{1}{2}, \frac{1}{2}} T_{1, \tilde{s}, J}^{\frac{1}{2}, \frac{1}{2}, \frac{1}{2}}\\
	\alpha_{2,2}^{T 2}
	&=B_{1,1}^2 B_{2,2}^0 N_{0, J, 0, J, J}^{2,2,2, S} \sum_{\tilde{s}=0}^1(-1)^{\frac{1}{2}+\tilde{s}-J} N_{2,1, \frac{1}{2}, \frac{1}{2}, \tilde{s}}^{1,1,\frac{1}{2}, \frac{1}{2}}
	T_{\tilde{s}, S, J}^{2,1, \frac{1}{2}} T_{1, \tilde{s}, J}^{\frac{1}{2}, \frac{1}{2}, \frac{1}{2}},\\
	\alpha_{2,3}^{T 2}
	&=B_{1,1}^2 B_{2,2}^0 N_{0, J, 0, J, J}^{2,2,2, S} \sum_{\tilde{s}=0}^1(-1)^{\frac{3}{2}-J} N_{2, S, \tilde{s}, \frac{1}{2}, J}^{1,1,1, \frac{1}{2}}
	T_{\frac{1}{2}, 1,\tilde{s}}^{1, \frac{1}{2}, \frac{1}{2}} T_{\tilde{s}, 1, J}^{\frac{1}{2}, \frac{1}{2}, \frac{1}{2}},\\
	\alpha_{3,1}^{T 2}
	&=\alpha_{2,1}^{T 2},\\
	\alpha_{3,2}^{T 2}
	&=\alpha_{2,3}^{T 2},\\
	\alpha_{3,3}^{T 2}
	&=B_{1,1}^2 B_{2,2}^0 N_{0, J, 0, J, J}^{2,2,2, S} \sum_{\tilde{s}=0}^1(-1)^{\frac{3}{2}-J} N_{2,1, \frac{1}{2}, \frac{1}{2}, \tilde{s}}^{1,1, \frac{1}{2},\frac{1}{2}}
	T_{\tilde{s}, S, J}^{2,1, \frac{1}{2}} T_{\tilde{s}, 1, J}^{\frac{1}{2}, \frac{1}{2}, \frac{1}{2}},
\end{align}
while analogues with $\mathbf{p} \rightarrow \mathbf{p}^{\prime}$, contributing to mixed matrix elements $\langle L=2|\hat{V}| L=0\rangle$, can be obtained from the above by Hermitian conjugation.
There is no tensor operator in the potential that can connect two $D$ waves. 

Other kinds of operators that only contribute to $P$ waves are
\begin{align}
	\big[\left(\mathbf{\hat{p}}^{\prime} \mathbf{\hat{p}}\right)_1\left(\sigma_1 \sigma_2\right)_1\big]_0 \big|J, \ell_d=0, L=1, S \big\rangle_\pi
	&=(4\pi)^2 \sum_{S^{\prime}} \alpha_{\pi, 1}^{V}\left(S, S^{\prime}\right)
	\big|J, \ell_d=0, L=1, S^{\prime} \big\rangle_{1} \,,
	\\
	\big[\left(\mathbf{\hat{p}}^{\prime} \mathbf{\hat{p}}\right)_1\left(\sigma_2 \sigma_3\right)_1\big]_0 \big|J, \ell_d=0, L=1, S \big\rangle_\pi
	&=(4\pi)^2 \sum_{S^{\prime}} \alpha_{\pi, 2}^{V}\left(S, S^{\prime}\right)
	\big|J, \ell_d=0, L=1, S^{\prime} \big\rangle_{1} \,,
	\\
	\big[\left(\mathbf{\hat{p}}^{\prime} \mathbf{\hat{p}}\right)_1\left(\sigma_1 \sigma_3\right)_1\big]_0 \big|J, \ell_d=0, L=1, S \big\rangle_\pi 
	&=(4\pi)^2 \sum_{S^{\prime}} \alpha_{\pi, 3}^{V}\left(S, S^{\prime}\right)
	\big|J, \ell_d=0, L=1, S^{\prime} \big\rangle_{1} \,,
\end{align}
with
\begin{align}
	\alpha_{1,1}^{V}\left(S, S^{\prime}\right)
	&=B_{1,1}^0 T_{1,0,1}^{1,1,1} N_{0, J, 1, S', J}^{1,1,1,S} T_{1, S, S'}^{2,1, \frac{1}{2}} N_{2,1, \frac{1}{2}, \frac{1}{2}, 1}^{1,1, \frac{1}{2},\frac{1}{2}} \,,
	\\
	\alpha_{1,2}^{V}\left(S, S^{\prime}\right)
	&=B_{1,1}^0 T_{1,0,1}^{1,1,1} N_{0, J, 1, S', J}^{1,1,1,S} N_{1, S, 1, \frac{1}{2}, S^{\prime}}^{1,1,1 \frac{1}{2}} T_{\frac{1}{2}, 1,1}^{1, \frac{1}{2}, \frac{1}{2}} \,,
	\\
	\alpha_{1,3}^{V}\left(S, S^{\prime}\right)
	&=-\alpha_{1,2}^{V}\left(S, S^{\prime}\right) \,,
	\\
	\alpha_{2,1}^{V}\left(S, S^{\prime}\right)
	&=B_{1,1}^0 T_{1,0,1}^{1,1,1} N_{0, J, 1, S', J}^{1,1,1,S} \sum_{\tilde{s}=0}^1(-1)^{\frac{3}{2}+\tilde{s}-S^{\prime}}
	N_{1, S, \tilde{s}, \frac{1}{2}, S^{\prime}}^{1,1,1, \frac{1}{2}} T_{\frac{1}{2}, 1, \tilde{s}}^{1, \frac{1}{2},\frac{1}{2}} T_{1, \tilde{s}, S^{\prime}}^{\frac{1}{2}, \frac{1}{2},\frac{1}{2}} \,,
	\\
	\alpha_{2,2}^{V}\left(S, S^{\prime}\right)
	&=B_{1,1}^0 T_{1,0,1}^{1,1,1} N_{0, J, 1, S', J}^{1,1,1,S} \sum_{\tilde{s}=0}^1(-1)^{\frac{1}{2}+\tilde{s}-S^{\prime}}
	N_{1,1, \frac{1}{2}, \frac{1}{2}, \tilde{s}}^{1,1, \frac{1}{2},\frac{1}{2}} T_{\tilde{s}, S, S^{\prime}}^{1,1, \frac{1}{2}} T_{1,\tilde{s}, S^{\prime}}^{\frac{1}{2}, \frac{1}{2}, \frac{1}{2}} \,,
	\\
	\alpha_{2,3}^{V}\left(S, S^{\prime}\right)
	&=B_{1,1}^0 T_{1,0,1}^{1,1,1} N_{0, J, 1, S', J}^{1,1,1,S} \sum_{\tilde{s}=0}^1(-1)^{\frac{3}{2}-S^{\prime}}
	N_{1, S,  \tilde{s}, \frac{1}{2}, S^{\prime}}^{1,1,1, \frac{1}{2}} T_{\frac{1}{2}, 1, \tilde{s}}^{1, \frac{1}{2}, \frac{1}{2}} T_{\tilde{s}, 1, S^{\prime}}^{\frac{1}{2}, \frac{1}{2}, \frac{1}{2}} \,,
	\\
	\alpha_{3,1}^{V}\left(S, S^{\prime}\right)
	&=B_{1,1}^0 T_{1,0,1}^{1,1,1} N_{0, J, 1, S', J}^{1,1,1,S} \sum_{\tilde{s}=0}^1(-1)^{\frac{1}{2}+\tilde{s}-S^{\prime}}
	N_{1, S, \tilde{s}, \frac{1}{2}, S^{\prime}}^{1,1,1, \frac{1}{2}} T_{\frac{1}{2}, 1, \tilde{s}}^{1, \frac{1}{2},\frac{1}{2}} T_{1, \tilde{s}, S^{\prime}}^{\frac{1}{2}, \frac{1}{2},\frac{1}{2}} \,,
	\\
	\alpha_{3,2}^{V}\left(S, S^{\prime}\right)
	&=-\alpha_{2,3}^{V}\left(S, S^{\prime}\right) \,,
	\\
	\alpha_{3,3}^{V}\left(S, S^{\prime}\right)
	&=B_{1,1}^0 T_{1,0,1}^{1,1,1} N_{0, J, 1, S', J}^{1,1,1,S} \sum_{\tilde{s}=0}^1(-1)^{\frac{3}{2}-S^{\prime}}
	N_{1,1, \frac{1}{2}, \frac{1}{2}, \tilde{s}}^{1,1, \frac{1}{2},\frac{1}{2}}, T_{\tilde{s}, S, S^{\prime}}^{1,1, \frac{1}{2}} T_{\tilde{s},1, S^{\prime}}^{\frac{1}{2}, \frac{1}{2}, \frac{1}{2}} \,.
\end{align}

The contribution of spin-orbit operators such as $O_7$ and $O_8$, which only contribute to $P$ waves, is
\begin{align}
	\big[\left(\mathbf{\hat{p}}^{\prime} \mathbf{\hat{p}}\right)_1 \sigma_1\big]_0 \big|J, \ell_d=0, L=1, S \big\rangle_\pi
	&=\frac{(4\pi)^2}{\sqrt{3}} \sum_{S^{\prime}} \alpha_{\pi, 1}^{L S}\left(S, S^{\prime}\right)\big|J, \ell_d=0,L=1, S^{\prime}\big\rangle_{1} \,,\\
	\big[\left(\mathbf{\hat{p}}^{\prime} \mathbf{\hat{p}}\right)_1 \sigma_2\big]_0 \big|J, \ell_d=0,L=1, S \big\rangle_\pi
	&=\frac{(4\pi)^2}{\sqrt{3}} \sum_{S^{\prime}} \alpha_{\pi, 2}^{L S}\left(S, S^{\prime}\right)\big|J, \ell_d=0,L=1, S'\rangle_{1} \,,\\
	\big[\left(\mathbf{\hat{p}}^{\prime} \mathbf{\hat{p}}\right)_1 \sigma_3\big]_0 \big|J, \ell_d=0,L=1, S \big\rangle_\pi
	&=\frac{(4\pi)^2}{\sqrt{3}} \sum_{S^{\prime}} \alpha_{\pi, 3}^{L S}\left(S, S^{\prime}\right)\big|J, \ell_d=0,L=1, S'\rangle_{1} \,,
\end{align}
with
\begin{align}
	\alpha_{1,1}^{L S}\left(S, S^{\prime}\right)
	&=N_{0, J, 1, S^{\prime}, J}^{1,1,1, S} T_{1,0,1}^{1,1,1} B_{1,1}^0 T_{1, S, S^{\prime}}^{1,1, \frac{1}{2}} T_{\frac{1}{2}, 1,1}^{1, \frac{1}{2}, \frac{1}{2}},
	\\
	\alpha_{1,2}^{L S}\left(S, S^{\prime}\right)
	&=\alpha_{1,1}^{L S}\left(S, S^{\prime}\right),
	\\
	\alpha_{1,3}^{L S}\left(S, S^{\prime}\right)
	&=N_{0, J, 1, S^{\prime}, J}^{1,1,1, S} T_{1,0,1}^{1,1,1} B_{1,1}^0
	(-1)^{1-S-S^{\prime}} T_{\frac{1}{2}, S, S^{\prime}}^{1, \frac{1}{2}, 1},
	\\
	\alpha_{2,1}^{L S}\left(S, S^{\prime}\right)
	&=N_{0, J, 1, S^{\prime}, J}^{1,1,1, S} T_{1,0,1}^{1,1,1} B_{1,1}^0
	(-1)^{\frac{3}{2}-S} T_{\frac{1}{2}, S, S^{\prime}}^{1, \frac{1}{2}, 1} T_{1,1, S^{\prime}}^{\frac{1}{2},\frac{1}{2}, , \frac{1}{2}},
	\\
	\alpha_{2,2}^{L S}\left(S, S^{\prime}\right)
	&=N_{0, J, 1, S^{\prime}, J}^{1,1,1, S} T_{1,0,1}^{1,1,1} B_{1,1}^0 \sum_{\tilde{s}=0}^1
	(-1)^{\frac{1}{2}+\tilde{s}-S^{\prime}} T_{\tilde{s}, S, S^{\prime}}^{1,1, \frac{1}{2}} T_{\frac{1}{2}, 1, \tilde{s}}^{1, \frac{1}{2}, \frac{1}{2}} T_{\tilde{s}, 1,S^{\prime}}^{\frac{1}{2}, \frac{1}{2}, \frac{1}{2}},\\
	\alpha_{2,3}^{L S}\left(S, S^{\prime}\right)
	&=N_{0, J, 1, S^{\prime}, J}^{1,1,1, S} T_{1,0,1}^{1,1,1} B_{1,1}^0 \sum_{\tilde{s}=0}^1
	(-1)^{\frac{3}{2}-S^{\prime}} T_{\tilde{s}, S, S^{\prime}}^{1,1, \frac{1}{2}} T_{\frac{1}{2}, 1, \tilde{s}}^{1, \frac{1}{2}, \frac{1}{2}} T_{1, \tilde{s}, S^{\prime}}^{\frac{1}{2}, \frac{1}{2}, \frac{1}{2}},\\
	\alpha_{3,1}^{L S}\left(S, S^{\prime}\right)
	&=\alpha_{2,2}^{L S}\left(S, S^{\prime}\right),\\
	\alpha_{3,2}^{L S}\left(S, S^{\prime}\right)
	&=\alpha_{2,1}^{L S}\left(S, S^{\prime}\right),\\
	\alpha_{3,3}^{L S}\left(S, S^{\prime}\right)
	&=\alpha_{2,3}^{L S}\left(S, S^{\prime}\right).
\end{align}

\begin{acknowledgments}
Y.C. and L.G. acknowledge the financial support of 
the European Union - Next Generation EU, Mission 4 Component 1, 
CUP F53D23001360006, for the PRIN 2022 project 
``Exploiting separation of scales in nuclear structure and dynamics''.
\end{acknowledgments}
	
\bibliography{bib.bib}
	
\end{document}